# Budget Optimization for Sponsored Search: Censored Learning in MDPs

**Kareem Amin** **Michael Kearns**
Computer and Information Science
University of Pennsylvania
{akareem,mkearns}@cis.upenn.edu

**Peter Key** **Anton Schwaighofer**
Microsoft Research
Cambridge, United Kingdom
{peter.key,antonsc}@microsoft.com

## Abstract

We consider the budget optimization problem faced by an advertiser participating in repeated sponsored search auctions, seeking to maximize the number of clicks attained under that budget. We cast the budget optimization problem as a Markov Decision Process (MDP) with *censored observations*, and propose a learning algorithm based on the well-known *Kaplan-Meier* or *product-limit* estimator. We validate the performance of this algorithm by comparing it to several others on a large set of search auction data from Microsoft adCenter, demonstrating fast convergence to optimal performance.

## 1 Introduction

In this paper we study algorithms for optimized budget expenditure in sponsored search. Given an advertiser's budget, the goal of such algorithms is to maximize the number of clicks obtained during each budgeting period. We consider a single-slot model in which our algorithm's competing bid — representing the rest of the "market" for clicks — is drawn from a fixed and unknown probability distribution. While in reality, advertisers (or their proxies) may often bid strategically and not stochastically, we view this assumption as analogous to classical models in finance, where despite strategic behavior of traders at the individual level, models of macroscopic price evolution that are stochastic (such as Brownian motion models) have been quite effective in developing both models and algorithms. Our empirical results will demonstrate that algorithms designed for these stochastic assumptions also perform quite well on the non-stochastic sequence of bids actually occurring in real search auctions.

The assumption of stochastic bids by the competing market leads to a Markov Decision Process (MDP) formulation of the optimal policy, where the states of the MDP specify the remaining time in the period and the remaining budget. However, the second-price nature of sponsored search introduces the challenge of *censored* observations: only if we win the click do we observe the actual competing price; otherwise, we only know our bid was too low.

Our main contributions are the introduction of efficient algorithms that combine the MDP formulation with the classical Kaplan-Meier [5] or product-limit estimator for censored observations, and a large-scale empirical demonstration that these algorithms are extremely effective in practice — even when our underlying distributional assumptions are badly violated. Our source of data is auction-level observations on hundreds of high-volume key-phrases from Microsoft adCenter. We show that our algorithms rapidly learn to compete with the strongest possible benchmark — the performance of an offline-optimal algorithm that knows the future competing bids, and always selects the cheapest clicks in each period.

## 2 Related Work

There is some prior work directly concerning the problem of optimizing an advertiser's budget [6, 12], as well as work concerned with characterizing the dynamics or equilibria of a market in which advertisers play from a family of optimizing strategies [2, 3]. All these works attempt to model the strategic behavior of agents participating in a sponsored search auction. We will depart from this, modeling the auction market stochastically, as is more common in finance.

The sponsored search budget optimization problem has also been formulated as an instance of online knapsack [12]. For the online knapsack problem, it is known that no online algorithm can converge to the optimum in the worst case [8]. The stochastic knapsack problem has also been studied, and there is an algorithm with near-optimal average-case performance [7]. One of our proposed algorithms is a censored learning version of such an algorithm. Our main proposed algorithm is closely modeled on an algorithm from a financial optimization problem [4], which similarly integrates a censored estimation step with greedy optimization.

Finally, we apply classical techniques from reinforcement learning in a finite state MDP, including Q-learning (c.f. [11]), as well as classical techniques from the study of censored observations [5, 9].

## 3 Preliminaries

The optimization problem we consider occurs over a series of *periods*. At the beginning of each period, the budget optimization algorithm is allocated a fresh budget $B$. This assumption is meant to reflect the manner in which advertisers actually specify their budgets in real sponsored search markets. Broadly speaking, the algorithm's goal is to maximize the number of clicks purchased, in each period, using the budget $B$.

Each period consists of a number of *auctions*, or an opportunity to earn a click. We consider the *single-slot* setting, wherein the search engine displays a single advertisement for each auction [1]. The algorithm places a *bid* in each auction. Before the bid is placed, a *price* is fixed by the auction mechanism. If the algorithm's bid exceeds this price, the algorithm wins an *impression*. For simplicity, we begin by assuming that winning an impression automatically guarantees that the algorithm also wins a click; we will describe later how to relax this assumption.

Once the algorithm wins a click, it is charged the price from its budget. Otherwise, the algorithm maintains its budget, and the next auction occurs. In actual sponsored search markets, the price is determined by the bids of arbitrary agents competing for the click in a modified second-price auction (see e.g. [10]).

The major assumption of this work is to instead model the prices as i.i.d. draws from an unknown distribution; we will refer to this price as the *market price*, since it represents the aggregate behavior of our algorithm's competitors. As in other financial applications, it is often analytically intractable to model the individual agents in a market strategically, so we instead consider the market stochastically as a whole. We will demonstrate that our algorithm outperforms other methods in practice on actual auction data from Microsoft adCenter, even when the i.i.d. assumption is badly violated.

We will consider some fixed, unknown, distribution $\mathcal{P}$ supported on $\mathbb{Z}^+$ with mass function $p(\cdot)$. On each auction, the market price is an independent random variable distributed as $\mathcal{P}$. We think of the budget $B$ and market prices as being expressed in terms of the smallest unit of currency that can be bid by the algorithm. Modeling the problem in this manner motivates a natural algorithm.

It is important to note that the market prices are not observed directly by the algorithm. Rather, the algorithm is only privy to the consequences of its bid (whether a click or impression is received), and changes to its budget.

Succinctly, we consider the following protocol:

**for** period $u = 1, 2, ...$ **do**
    $B_{u,T} = B$
    **for** auctions remaining $t = T, T-1, ..., 1$ **do**
        Algorithm bids $b_{u,t} \leq B_{u,t}$.
        Nature draws price $x_{u,t} \sim \mathcal{P}$.
        **if** $b_{u,t} \geq x_{u,t}$ **then**
            $c_{u,t} \leftarrow 1$
            $B_{u,t-1} \leftarrow B_{u,t} - x_{u,t}$
        **else**
            $c_{u,t} \leftarrow 0$
            $B_{u,t-1} \leftarrow B_{u,t}$
        **end if**
        Algorithm observes $c_{u,t}$, $B_{u,t-1}$
    **end for**
**end for**

When an algorithm places a large enough a bid, winning the click, it also observes the true market price, since its budget it reduced by that amount. However, should the algorithm fail to win the click, it only knows that the market bid was higher than the bid placed. Thus, the algorithm receives what is known in the statistical literature as partially *right-censored observations* of the market prices $\{x_{u,t}\}$.

Informally, we always assume that an algorithm has available to it any information it would have in a real sponsored search auction (although not always at the same granularity), and no more. So it may be informed of whether it received a click or impression on an auction-by-auction basis, but not information regarding prices if it did not win the click.

Finally, we assume there is only a single keyword which the advertiser is bidding on. Our methods generalize to the setting where there are multiple keywords with multiple click-through rates and valuations for a click. However, for simplicity, we do not consider these extensions in this work.

[1] We suspect our methods can be adapted to the multi-slot case, but leave it to future work.

### 3.1 Notation

Given a distribution $\mathcal{P}$, supported on $\mathbb{Z}^+$, with mass function $p$, we let the tail function $T_p(b) = \sum_{b'=b+1}^{\infty} p(b')$ denote the mass to the right of $b$.

We use $[N]$ to mean the set $\{1, ..., N\}$, and $[N]_0 = [N] \cup \{0\}$.

## 4 MDP Formulation

An algorithm for the optimization problem introduced in the previous section can be described as an agent in a Markov Decision Process (MDP). An MDP $\mathcal{M}$ can be written as $\mathcal{M} = (\mathcal{S}, \{A_s\}_{s\in\mathcal{S}}, \mu, r)$ where $\mathcal{S}$ is a set of states, and $A_s$ are the set of actions available to the agent in each state $s$, and $\mathcal{A} = \cup_{s\in\mathcal{S}} A_s$. For $a \in A_s$, $\mu(a, s, s')$ is the probability of transitioning from state $s$ to state $s'$ when taking action $a$ in state $s$. $r(a, s, s')$ is the expected reward received after taking action $a$ in state $s$ and transitioning to state $s'$. The goal of the agent is to maximize the expected reward received while transitioning through the MDP.

In our case, the state space is given by $\mathcal{S} = [B]_0 \times [T]_0$. With $t$ auctions remaining in period $u$, the algorithm is in state $(B_{u,t}, t) \in \mathcal{S}$. Furthermore, the actions available to any algorithm in such a state are the set of bids that are at most $B_{u,t}$. So for any $(b, t) \in \mathcal{S}$ we let $\mathcal{A}_{(b,t)} = [b]_0$.

When $t \geq 1$, two types of transitions are possible. The agent can transition from $(b, t)$ to $(b, t - 1)$ or from $(b, t)$ to $(b', t-1)$ where $0 \leq b' < b$. In the former case, the agent must place a bid lower than the market price. Therefore, we have that $\mu(a, (b, t), (b, t - 1)) = T_p(a)$. Furthermore, the agent does not win a click in this case, and $r(a, (b, t), (b, t - 1)) = 0$. In the latter case, let $\delta = b - b'$. The agent must bid at least $\delta$, and the market price must be exactly $\delta$ on auction $t$ of the period in question. Therefore, we have that $\mu(a, (b, t), (b', t-1)) = p(\delta)$ and $r(a, (b, t), (b', t-1)) = 1$ so long as $a \geq \delta$.

When $t = 0$, for any action, the agent simply transitions to $(B, T)$ with probability 1, with no reward. The agent's budget is refreshed, and the next period begins.

All other choices for $(a, s, s') \in \mathcal{A}_s \times \mathcal{S} \times \mathcal{S}$ represent invalid moves, and hence $\mu(a, s, s') = r(a, s, s') = 0$.

Finally, conditioned on an agent's choice of action $a$ and current state $s$, its next state $s$ is independent of all previous actions and states, since the market prices $\{x_{u,t}\}$ are independent. The Markov property is satisfied, and we indeed have an MDP.

We call this the *Sponsored Search MDP* (SS-MDP). If $\pi$ is a fixed mapping from $\mathcal{S}$ to $\mathcal{A}$ satisfying $\pi(s) \in \mathcal{A}_s$, we say that $\pi$ is a *policy* for the MDP.

Note that an agent in the SS-MDP, started in an arbitrary state $(b, t)$, arrives at the state $(B, T)$ after exactly $t + 1$ actions. Therefore, we can define the random variable $C_\pi(b, t)$ to be the total reward (i.e. number of clicks) attained by policy $\pi$ before returning to $(B, T)$. We say that $\pi$ is *an optimal policy for the SS-MDP* if an agent started at $(B, T)$, playing $\mu(s)$ in each state $s$ encountered, maximizes the expected number of clicks rewarded before returning to $(B, T)$. In other words:

**Definition 1.** *A policy $\pi^*$ is an optimal policy for the SS-MDP if $\pi^* \in \arg\max_\pi E[C_\pi(B, T)]$.*

The SS-MDP is determined by the choice of budget $B$, time $T$ and distribution $p$. We will want to make the optimal policy's dependence on $p$ explicit, and consequently we will write it a $\pi_p^*$.

## 5 The Value Function

MDPs lend themselves to dynamic programming. Indeed, we can characterize exactly the optimal policy for the SS-MDP when the probability mass function $p$ is known.

For a distribution $p$, let $V_p(b, t)$, the *value function for* $p$, denote the expected number of clicks received by an optimal policy started at state $(b, t)$. That is, if $\pi_p^*$ is an optimal policy, then $V_p(b, t) = E[C_{\pi^*}(b, t)]$.

First note that when $T = 0$, $V_p(B, T) \equiv 0$. Consider the policy $\pi^*_{a,b,t}$ that takes action $a$ in state $(b, t)$, and plays optimally thereafter. Let $V_p(a, b, t) = E[C_{\pi^*_{a,b,t}}(b, t)]$ be the clicks received by such a policy from state $(b, t)$. Observe that $V_p(a, b, t)$ can be written in terms of $V_p(\cdot, t - 1)$:

$$V_p(a, b, t) = \sum_{\delta=1}^{a} p(\delta)[1 + V_p(b - \delta, t - 1)] + T_p(a)V_p(b, t - 1).$$

In other words, if the market price is $\delta \leq a$, which occurs with probability $p(\delta)$, the agent will win a click at the price of $\delta$ and transition to state $(b - \delta, t - 1)$. At this point it behaves optimally, earning $V_p(b - \delta, t - 1)$ clicks in expectation. If the market price is greater than $a$, which occurs with probability $T_p(a)$, the agent will retain its budget, transitioning to the state $(b, t-1)$, earning $V_p(b, t-1)$ clicks in expectation. Furthermore, we know that:

$$V_p(b, t) = \arg\max_{a \leq b} V_p(a, b, t).$$

Therefore, if $p$ and $V_p(\cdot, T - 1)$ are known then we can compute $V_p(a, b, t)$ in $O(B)$ operations, and so compute $V_p(b,t)$ in $O(B)$ operations. Recalling that $V_p(b,0) \equiv 0$, we can compute $V_p(b,t)$ for all $(b,t) \in [B]_0 \times [T]_0$ in $O(B^2T)$ operations.

# 6 Censored Data

In the previous sections, we described how to compute the optimal policy $\pi^*_p$ when $p$ is known. A natural algorithm for budget optimization is therefore to maintain an estimate $\hat{p}$ of $p$, and bid greedily according to $\pi^*_{\hat{p}}$. Before describing such an algorithm, we will discuss the problem of estimating $p$.

As introduced in Section 3, the observations received by a budget-optimization algorithm are partially right-censored data. We begin with a general discussion of censoring.

Suppose that $\mathcal{P}$ is a distribution with mass function $p$ and $(z_1, ..., z_n)$ are i.i.d., $\mathcal{P}$-distributed random variables. Fix $n$ integers $k_1, ..., k_n$, and define $o_i = \min(z_i, k_i)$. We say that the sample $\{o_i\}$ is partially right-censored data.

If $o_i < k_i$, we say that $o_i$ is a direct observation. In other words, $o_i = z_i$ and we have observed the true value of $z_i$. Otherwise, $o_i = k_i$ and we say that $o_i$ is a censored observation. We know only that $z_i \geq k_i$.

Given such partially right-censored data, the Product-Limit estimator [5] is the non-parametric maximum-likelihood estimator for $p$.

**Definition 2.** *Let $\mathcal{P}$ be a discrete distribution with mass function $p$, and let $\{z_i\}$ be i.i.d. $\mathcal{P}$-distributed random variables. Given integers $K = (k_1, ..., k_n)$ and observations $O = (o_1, ..., o_n)$ where $o_i = \min(z_i, k_i)$, let $PL(K,O)$ be the Product-Limit estimator for $p$.*

Specifically, given integers $K$, and a set of observations $O$ generated by a distribution $\mathcal{P}$, let $D(s) = |\{o_i \in O \mid s = o_i < k_i\}|$ be the number of direct observations of value $s$, and $N(s) = |\{o_i \in O \mid s \leq o_i, s < k_i\}|$. Now let $S(t) = \prod_{s=1}^{t-1} 1 - \frac{D(s)}{N(s)}$. The CDF of $PL(K,O)$ is given by $1 - S(t)$.

In our setting, we are receiving censored observation of the random variables $\{x_{u,t}\}$ where the censoring set $K$ is given by $\{b_{u,t}+1\}$, and $o_{u,t} = \min\{b_{u,t}+1, x_{u,t}\}$. When a click is received (i.e. $x_{u,t} < b_{u,t}+1$), we observe $x_{u,t}$ directly since $x_{u,t} = B_{u,t} - B_{u,t-1}$, the amount which the algorithm is charged for winning the click. Otherwise, the algorithm is charged nothing, and we only know that $x_{u,t} \geq b_{u,t}+1$.

Finally, we will eventually consider the setting in which winning an impression does not necessarily guarantee winning a click. In such a setting, we will have both left-censored and right-censored observations of $x_{u,t}$, what is known as *doubly-censored data*.

If the algorithm bids $b_{u,t}$ and does not win the impression, we know that $x_{u,t} \geq b_{u,t}+1$. Similarly, if the algorithm wins both the impression and the click, it gets to observe $x_{u,t}$ directly. However, should the algorithm win the impression but not the click, it is only informed that it placed a large-enough bid (that it won the impression), or $x_{u,t} < b_{u,t}+1$, without observing $x_{u,t}$ directly. We give a more detailed discussion of this setting in Section 9. For doubly-censored data, there is algorithm giving the non-parametric MLE [9].

# 7 Greedy Product-Limit Algorithm

The algorithm we propose maintains an estimate $\hat{p}$ of $p$. With budget $b$ remaining, and $t$ auctions remaining, the algorithm will greedily use its current estimate of $p$, and bid $\pi_{\hat{p}}(b,t)$. The pseudo-code for *Greedy Product-Limit* contains a detailed description.

**Algorithm 1** Greedy Product-Limit

**Input:** Budget $B$

Initialize distribution $\hat{p}$ uniform on $[B]$<br>
Initialize $K = []$; Initialize $O = []$<br>
**for** period $u = 1, 2, ...$ **do**<br>
    Set $B_{u,T} := B$<br>
    **for** auctions remaining t=T,T-1,...,1 **do**<br>
        Bid $\pi^*_{\hat{p}}(B_{u,t}, t)$<br>
        Set $k_{u,t} \leftarrow \pi^*_{\hat{p}}(B_{u,t}, t) + 1$<br>
        $K \leftarrow [K, k_{u,t}]$<br>
        **if** Click won at price $x_{u,t}$ **then**<br>
            $O \leftarrow [O, x_{u,t}]$<br>
        **else**<br>
            $O \leftarrow [O, k_{u,t}]$<br>
        **end if**<br>
        Update $\hat{p}$ to $PL(K,O)$<br>
    **end for**<br>
**end for**

# 8 Competing Algorithms

In this section we will describe a few alternative strategies against which we compare *Greedy Product-Limit*. The first relies on an observation that, given an *arbitrary* sequence of market prices, there is a simple bidding strategy that has a constant competitive ratio to the offline optimal.

## 8.1 Offline Optimality

So far we have focused our attention on the notion of optimality introduced in Section 4. Namely, an algorithm is optimal if it achieves $V_p(B,T)$ clicks, in expectation, in every period. However, given an arbitrary (non-stochastic) vector of $T$ market prices $\mathbf{x} = (x_1, ..., x_T)$, we can define $C^*(\mathbf{x}, B)$ to be the maximum number of clicks that could be attained by any sequence of bids, knowing $\mathbf{x}$ *a priori*. In other words, if $\mathbf{b} \in \{0,1\}^T$, and $\|\mathbf{b}\|_0 = |\{b_i \mid b_i = 1\}|$, then we define

**Definition 3.**

$$C^*(\mathbf{x}, B) \triangleq \max_{\mathbf{b}\in\{0,1\}^T} \|\mathbf{b}\|_0 \textit{ subject to } \mathbf{x} \cdot \mathbf{b} \leq B.$$

We call a sequence of bids for $\mathbf{x}$ that attains $C(\mathbf{x}, B)$ clicks an *optimal offline policy*. Notice that one attains the optimal offline policy by greedily selecting to win the clicks with the cheapest prices, until the budget $B$ is saturated.

### 8.2 Fixed Price

Competing against the notion of optimality introduced in Section 8.1 may seem onerous in the online setting. Indeed, competing against an arbitrary sequence of prices is a special case of the online knapsack problem, which is known to be hard. However, we will now show that for any sequence of prices $\mathbf{x}$, there always exists a simplistic bidding policy which would have attained a constant factor of the bids of the optimal offline policy.

Let $Fixed(b)$ be the policy that bids $b$ on every auction that it has budget to do so, and define $C(\mathbf{x}, b, B)$ to be the number of clicks attained by $Fixed(b)$ against $\mathbf{x}$ with budget $B$.

**Theorem 1.** *For any sequence of prices $\mathbf{x}$, and budget $B$, there exists a bid $b$ such that $C(\mathbf{x}, b, B) \geq \frac{1}{2}C^*(\mathbf{x}, B)$.*

*Proof.* Let $b^*$ be the value of the price for the most expensive click that the optimal offline policy selects to win. Suppose that the optimal offline policy wins $M + N$ clicks, where $M$ clicks were won with a price of exactly $b^*$ and the remaining $N$ clicks were won with a price of $b^* - 1$ or less.

If $N \geq \frac{1}{2}C^*(\mathbf{x}, B)$, then $Fixed(b^* - 1)$ would win all $N$ clicks, giving the desired result.

Otherwise, we know that $M \geq \frac{1}{2}C^*(\mathbf{x}, B)$. Consider the policy $Fixed(b^*)$. In the worst case, the policy will win only clicks with price $b^*$ before saturating its budget. However, we know that $Mb^* \leq B$, and so $C(\mathbf{x}, b^*, B) \geq M \geq \frac{1}{2}C^*(\mathbf{x}, B)$, as desired.

□

### 8.3 Fixed-Price Search

This motivates a simple algorithm which attempts to find the best fixed-price, *Fixed-Price Search*.

**Algorithm 2** Fixed-Price Search

Select $b_1$ arbitrarily.<br>
**for** period $u = 1, 2, ...$ **do**<br>
    $C_u := 0$<br>
    **for** auctions remaining $t = T, T-1, ..., 1$ **do**<br>
        Bid $b_u$<br>
        **if** Click won **then**<br>
            $C_u \leftarrow C_u + 1$<br>
        **end if**<br>
    **end for**<br>
    $b_{u+1} \leftarrow UpdateBid(\{b_{u'}, C_{u'}\}_{u'=1}^{u})$<br>
**end for**

The algorithm plays a fixed-price strategy each period. At the end of the period it uses a subroutine *UpdateBid* to select a new fixed-price according to how many clicks it has received. There are many reasonable ways to specify the *UpdateBid* subroutine, including using additive or multiplicative updates (e.g. treating each price as an expert and running a bandit algorithm such as Exp3 [1]). But general, the performance of any such approach cannot overcome the fixed-price "gap" of $E_{\mathbf{x}}[C^*(\mathbf{x}, B) - \max_b C(\mathbf{x}, b, B)]$.

### 8.4 Q-learning

Given the MDP formulation of the problem in Section 4, we may hope to solve the problem using techniques from reinforcement learning. Q-learning with exploration is one of the simplest algorithms for reinforcement learning, giving good results in a number of applications.

In Q-learning, the agent begins with an estimate $Q(a, b, t)$ of the function $V_p(a, b, t)$, called the Q-value, for each state $(b, t)$ and action $a \in [b]$. In a state $(b, t)$, the agent greedily performs the best action $a^*$ for that state using the current Q-values receiving some reward $\hat{r}$ and arriving at a new state $(b', t-1)$. The Q-values for state $(b', t-1)$ and the observed reward $\hat{r}$ are then used to update $Q(a^*, b, t)$. This is often combined with forced exploration. Notice that Q-learning will necessarily ignore the special assumptions placed on the underlying MDP. In particular, from our discussion in Section 4, we have that $\pi(a, (b_1, t_1), (b'_1, t_1 - 1)) = \pi(a, (b_2, t_2), (b'_2, t_2 - 1))$ when $b_1 - b'_1 = b_2 - b'_2$.

### 8.5 Knapsack Approaches

As referenced in Section 2, the problem of budget optimization in sponsored search is very related to the online knapsack problem. In the online knapsack problem, an optimizer is presented with a sequence of items with values and weights. At each time step, the optimizer makes an irrevocable decision to take the item (subtracting its weight from the optimizer's budget, and gaining its value). In the worst case, Marchetti-Spaccamela et al. demonstrate that a constant-factor competitive ratio with the offline is not possible [8]. Nevertheless, there are many results from the online knapsack literature that are applicable to our setting.

While in the worst case the online knapsack problem is hard, Lueker gives an average-case analysis for an algorithm for the *Stochastic Knapsack Problem*, which is related to our setting [7]. In the Stochastic Knapsack Problem, items $\{(r_i, x_i)\}$ are i.i.d. draws from some fixed, *known*, distribution. $r_i$ is the profit or reward earned by taking the item, and $x_i$ is the price of the item. In our setting, all clicks are considered indistinguishable for a fixed keyword, and so $r_i = 1$.

Note that the protocol differs from ours in a few ways. Firstly, there is no learning. The underlying distribution is assumed to be known. Secondly, there is no censoring of data, or any notion of an auction. The optimizer is presented with each item up-front, at which point it must make a decision before moving on to the next. Thirdly, in the language of our setting, there is only a single period.

Under these assumptions, the algorithm of Lueker gives a simple algorithm which differs from the true optimum by an average of $\Theta(T)$, where $T$ is the length of the period, assuming that the budget available scales with $T$ [7].

Nevertheless, the same ideas behind Greedy Product-Limit give us a natural adaptation of this algorithm to our setting. Suppose that the prices are presented up-front and that $\mathcal{P}$ is known. With budget $B$ remaining and time $T$ remaining in a period, the algorithm computes:

$$v(B/T) \triangleq \max_v \{v \mid \sum_{a=1}^{v} a \cdot p(a) \leq B/T\}$$

and takes the click iff its market price $x$ satisfies $x \leq v(B/T)$. Thus, when the prices are not presented up-front, it is equivalent to simply bidding $v(B/T)$.

Note that bidding $v(B/T)$ is natural; it is the bid that, in expectation, costs $B/T$, or smooths the remaining budget over the time remaining. We can now combine this with an estimation step, using the product-limit estimator, as we did for *Greedy Product-Limit*, simply replacing line 7 with the assignment $k_{u,t} := v(B/T)$. We refer to this strategy as *LuekerLearn*.

Notice, however, that once $\hat{p}$ has converged to the true $p$, we should not expect this algorithm to outperform *Greedy Product-Limit*, which would bid optimally. For a fixed choice of distribution $p$, budget $B$, and number of auctions $T$, let $L_p(B, T)$ be the expected number of clicks earned by running the algorithm of Lueker, knowing $p$. $L_p(B, T) \leq V_p(B, T)$, by definition of $V_p(B, T)$. We will comment (as established by Lueker) that the gap $V_p(B, T) - L_p(B, T)$ is exacerbated by distributions with large variance relative to $B$ and $T$, an issue we will return to in Section 9; see also Figure 1.

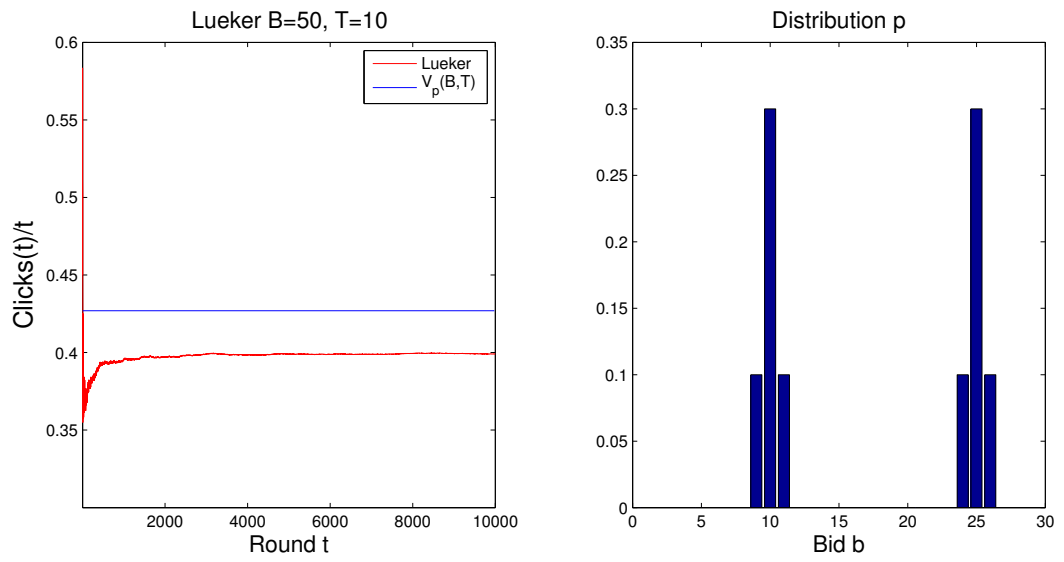

Figure 1: A distribution for which $L_p(B, T)$ is bounded away from $V_p(B, T)$. The red curve plots the performance (averaged over auctions) of the algorithm of Lueker's algorithm when the distribution $p$ is known. The distribution is displayed on the right.

### 8.6 Budget Smoothing

There is also literature in which other budget-smoothing approaches are considered. Zhou et al. consider the budget optimization problem in sponsored search as a online knapsack problem directly [12]. In our setting, their algorithm guarantees a $\ln(B) + 1$ competitive ration with the offline optimum. Their algorithm smooths its budget over time, and operates by bidding: $1/(1+\exp(z(t)-1))$ where $z(t)$ is the fraction of budget remaining at time $t$.

## 9 Experimental Results

In this section we will describe experimental results for the previously described algorithms. We use bids placed through Microsoft's adCenter in two sets of experiments. In the first, we assume that our modeling assumptions from Section 3 are correct, and construct a distribution $\mathcal{P}$ from the empirical data for use in simulation. In the second set of experiments, we run the methods on the historical data directly, taken as an individual sequence and thus violating our stochastic assumptions. We will see that in both cases, our suggested algorithm outperforms the other methods discussed. First, however, we discuss an important generalization to the setting that we have considered so far.

### 9.1 Impressions and Clicks

Until now we have assumed that all ad impressions result in a click (i.e. winning an auction results in an automatic click). We will now relax this assumption. Instead, when an advertiser wins an impression, we will suppose that whether a click occurs is an independent Bernoulli random variable with mean $r$. We call $r$ the *click-through rate*. If a click does indeed occur, the advertiser is charged the market price. Otherwise, the advertiser is informed that an impression has occurred, but maintains its budget.

All the methods described generalize to this setting in a straightforward manner. Nevertheless, it is worth being explicit about how *Greedy Product-Limit* must be modified. First note that the MDP formulation for the problem differs in the definition of the transition probability $\mu$. In particular, we now have $\mu(a,(b,t),(b-\delta,t-1)) = rp(\delta)$ (when $\delta \leq a$), and $\mu(a,(b,t),(b,t-1)) = (1 - r\sum_{\delta=1}^{a} p(\delta))$. $\pi_p^*$ can still be computed using dynamic programming, where:

$V_p(a,B,T) = (1 - r\sum_{\delta=1}^{a} p(\delta))V_p(B,T-1) + \sum_{\delta=1}^{a} rp(\delta)[1 + V_p(B-\delta,T-1)]$

and $\pi_p^*(B,T) = \arg\max_{a \leq B} V_p(a,B,T)$

Furthermore, as discussed in Section 6, rather than using the Product-Limit estimator, this setting requires that the new algorithm treat doubly-censored data, for which techniques exist [9].

### 9.2 Data

The data used for these experiments were generated by collecting the auction history from advertisers placing bids through Microsoft's adCenter over a six month period. For a given keyword, we let the number of times that keyword generated an auction be its *search volume*, $\texttt{Vol}_k$, and take keyword $k$ to be the keyword with the $k$-th largest volume.

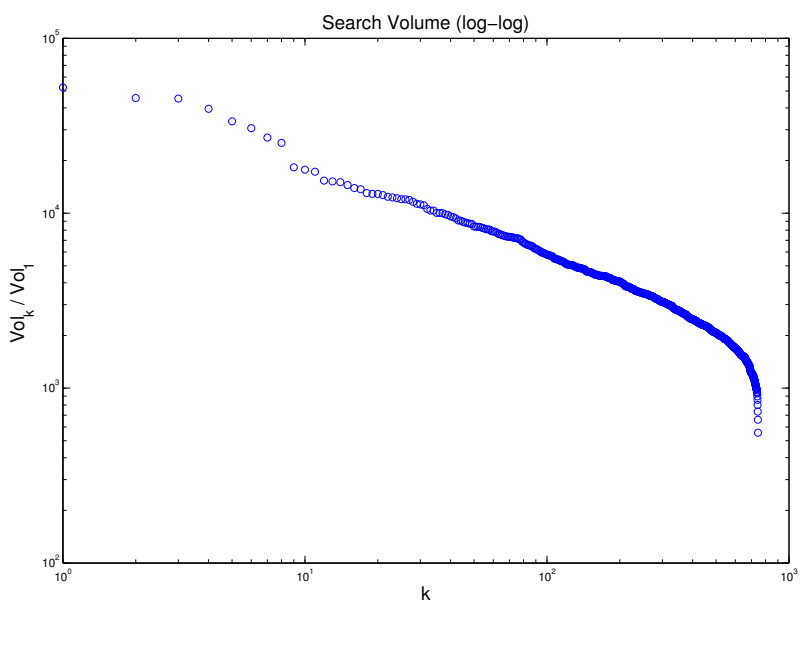

Figure 2: Log-log plot of $\texttt{Vol}_k/\texttt{Vol}_1$ for each $k$

The distribution of search volume is clearly heavy-tailed (see Figure 2) and is well approximated by a power law over several orders of magnitude.

We ran our experiments on the 100 keywords with largest search volume. As we will discuss further in the next section, the bidding behavior is quite varied among the different keywords in the data set.

In actual sponsored search-auctions, each bidder is given a *quality-score* for each keyword. Bidders' actual bids are multiplied by these quality scores to determine who wins the auction. For each keyword $k$, and auction $t$, let $x_{k,t}$ be the *quality-score-adjusted bid* for the advertiser that historically won the top slot for that auction, and $c_{k,t}$ indicate whether a click occurred.

We take the perspective of a new advertiser with unit quality score. $x_{k,t}$ is the amount that such an advertiser would have needed to bid to have instead taken the top slot (and the amount the advertiser would be charged should they also receive a click).

Finally, we make the single-slot assumption throughout, so even if multiple ads were indeed shown historically, we assume that an algorithm wins an impression if and only if it wins the top slot. In principle, *Greedy Product-Limit* can be modified to operate when multiple ad slots are available. However, we avoid this complication in this work.

### 9.3 Distributional Simulations

The first set of simulations use the historical data $\{x_{k,t}\}$, to construct an empirical distribution $p_k$ on market prices for each keyword $k$. We also set a fixed click-through rate $r_k$, for each keyword using the background click-through rate for that keyword (the empirical average of $\{c_{k,t}\}$). While our main results in the next section eliminate the distribution $p_k$ and use the sequence $\{x_{k,t}\}$ directly, we first simulate and investigate the case where our modeling assumptions hold.

Figure 3 demonstrates that the types of distributions generated in this manner are quite varied.

The budget $B_k$ allocated to the advertiser for keyword $k$ is selected so that, optimally, a constant fraction of clicks are available. In other words, the simulations were run with $B_k(T)$ satisfying $V_{p_k}(B_k(T),T) = fT$, where $f = 10\%$. Each experiment was run for 10 periods containing $T = 100$ auctions each, and 20 experiments were run for each keyword.

A major observation is that *Greedy Product-Limit* converges to the optimum policy on the time-scale of auctions, not periods. This is significant since certain methods considered are doomed to converge on the time-scale of periods instead. For example, each state can be visited at most once by Q-learning in a single period. Furthermore, for a particular time $t$, the only state corresponding to that $t$ visited is $(B_{u,t}, t)$, (i.e. the state corresponding to the budget held by the algorithm at that time). Similarly, the Fixed-Price algorithm adjusts its bid at the end of every period. Table 1 shows that after just 2 periods, *Greedy Product-Limit* (GPL) has come within five percent of optimal across all keywords. For each algorithm, the results are the averages of 20 different simulations, and are statistically significant. An unpaired 2-sample t-test between the results for GPL and those of any other algorithm yields p-values that are less than $10^{-20}$.

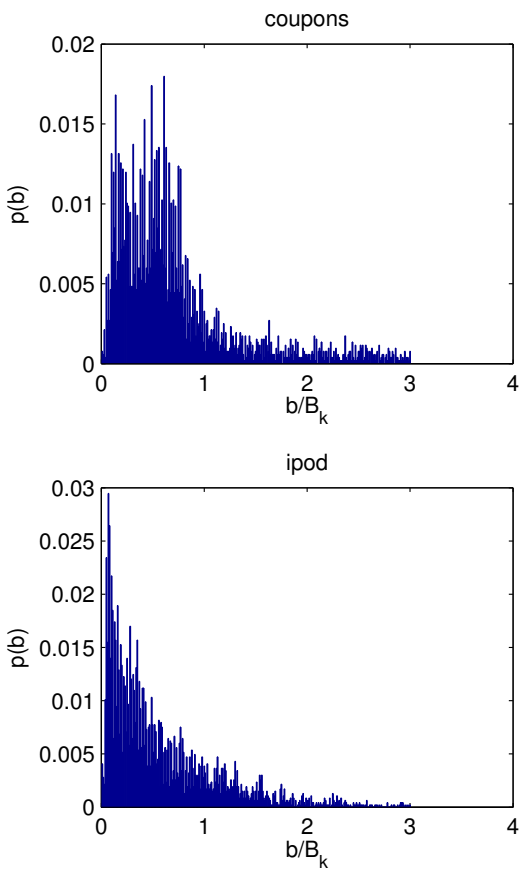

Figure 3: Each plot represents the empirical distribution $p_k$ for a different keyword $k$. The x-axis represents the bid as a fraction of the total budget $B_k$ allocated to the advertiser for keyword $k$.

Table 1: Average Competitive Ratio with $V_{p_k}(B,T)$, across all keywords and experiments, after two periods.

| Algorithm Name | Competitive Ratio | Std |
|---|---|---|
| Greedy Product-Limit | 0.9573 | 0.1704 |
| LuekerLearn | 0.8448 | 0.1842 |
| Fixed-Price Search | 0.8352 | 0.1733 |
| Q-learn | 0.7484 | 0.1786 |
| Budget Smoothing | 0.1597 | 0.2418 |

### 9.4 Sequential Experiments

The main result of our work comes from experiments run on the real sequential data $\mathbf{d}_k = \{x_{k,t}, c_{k,t}\}_t$. Rather than taking $\{x_{k,t}, c_{k,t}\}$, and constructing the distribution $p_k$ and a click-through rate, as in the previous section, we can use the sequence directly. Each of the previous methods are well-defined if the prices and clicks are generated in this manner, as opposed to being generated by the stochastic assumptions that motivated *Greedy Product-Limit*. We break the sequence into 10 periods of length $T = 100$. In reality, the number of auctions in a period might vary. However, this is minor, as an advertiser can attain good estimates of period-length. We allocate to each algorithm the same budget $B_k$ used in the stochastic experiments.

Recall the notion of offline optimality defined in section 8.1. For each keyword $k$, we can compute the exact auctions that one should win knowing the sequence $\{x_{k,t}, c_{k,t}\}$ *a priori*. We will demonstrate that *Greedy Product-Limit* is competitive with even this strong notion.

We first look at the nature and time-scale of the convergence of *Greedy Product-Limit*.

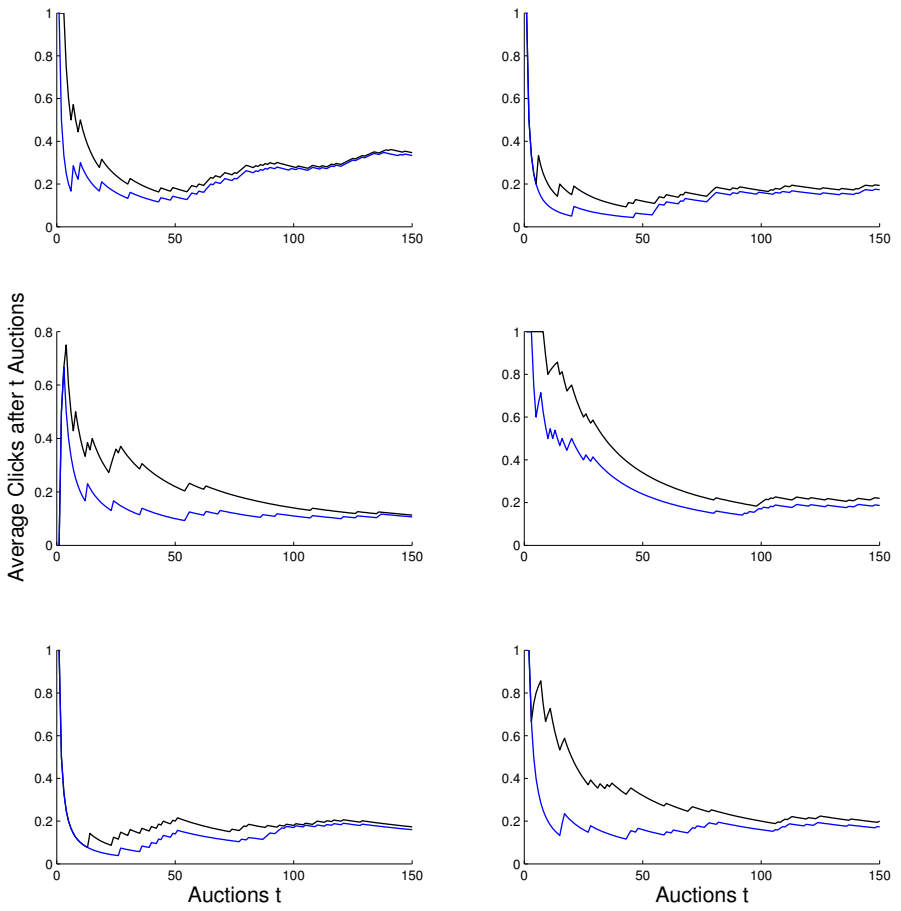

Figure 4: Convergence rates for 6 different keywords. The x-axis denotes auctions $t$, and the y-axis plots, in black, the number of clicks attained by the offline optimal after $t$ auctions, normalized by $t$. The blue plot shows the same for *Greedy Product-Limit*. *Greedy Product-Limit* converges in the auction time-scale, not the period time-scale.

We will shortly see that *LuekerLearn*, the modification of *Greedy Product-Limit* attains similar performance. Like *Greedy Product-Limit*, it converges in the time-scale of auctions. However, recalling Figure 1, *LuekerLearn* will sometimes converge to something suboptimal, especially on keywords where there is a lot of variance in the bids. Figure 5 demonstrates this behavior on the sequential data. In fact, let $A_k$ denote the cumulative number of clicks attained by *Greedy Product-Limit* and $L_k$ denote the cumulative number of clicks attained by *LuekerLearn* after 10 periods for keyword $k$, defining $Z_k = A_k/L_k$ and $S_k = std(\{x_{k,t}\}_t)$. The observations $\{Z_k\}$ are positively correlated with $\{S_k\}$, with a correlation coefficient of 0.2103 that is significant with a $p$-value of 0.0357.

Figure 6 demonstrates that, ignoring variance, *Greedy Product-Limit* has indeed converged to optimal across all keywords after just a single period. Let $O_k$ be the offline optimal number of clicks that can be attained for keyword $k$ after 10 periods (the entire data set). Let $A_{k,p}$ denote the cumulative number of clicks attained by an algorithm on keyword $k$ after $p$ periods. For an algorithm that is optimal after a single period, we should expect $A_{k,p}/O_k$ to be roughly $p/10$ for each period $p$. Indeed, while there are certain keywords for which this doesn't happen, we see that this is true on average.

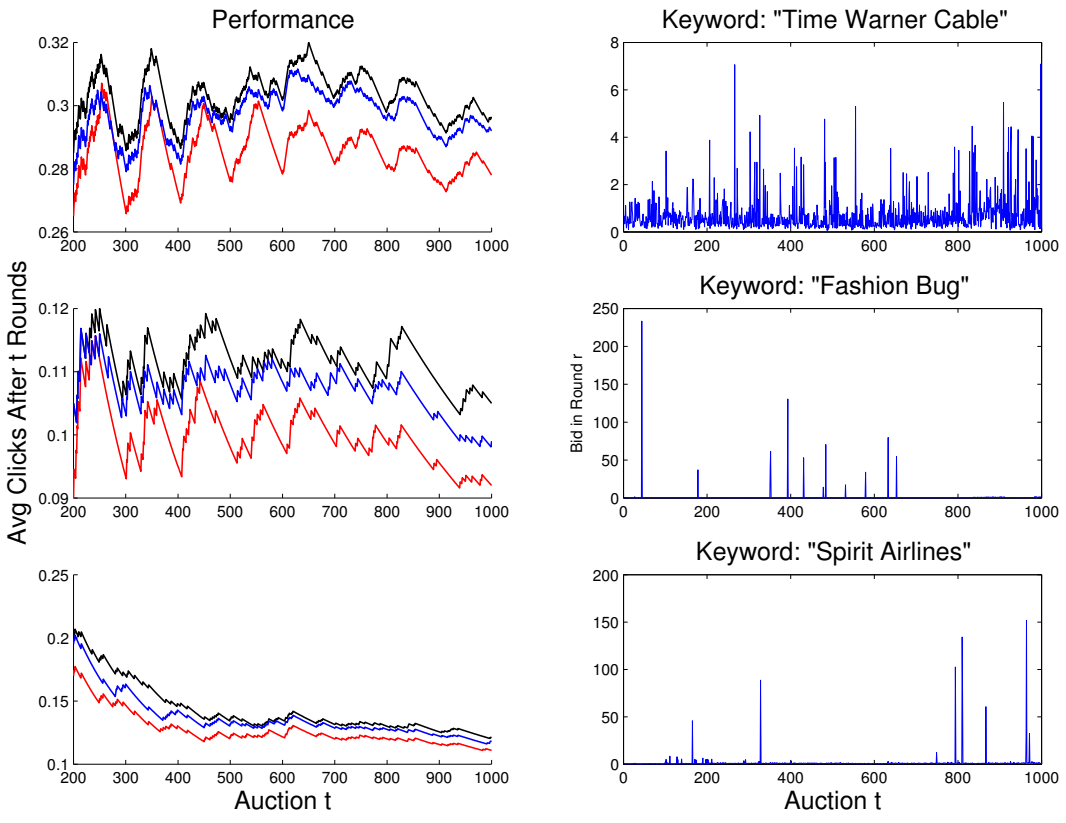

Figure 5: On the left the y-axis plots, in black, the number of clicks attained by the offline policy after $t$ auctions, normalized by $t$. The blue plot shows the same for *Greedy Product-Limit* and the red for *LuekerLearn*. In all three, *LuekerLearn* is bounded away from the offline optimal. The right side displays $x_{k,t}/\bar{x}_k$ where $\bar{x}_k$ is the average over $x_{k,t}$ for the corresponding keyword. Note the "bursty" nature of the market price, with auctions occurring that set a market price hundreds of times greater than the average.

Figure 7 and Table 2 summarize the performance of the competing methods.

Table 2: Average Competitive Ratio with $O_K$, across all keywords.

| Algorithm Name | Competitive Ratio | Std |
|---|---|---|
| Greedy Product-Limit | 0.9062 | 0.1166 |
| LuekerLearn | 0.8962 | 0.1152 |
| Fixed-Price Search | 0.6253 | 0.1395 |
| Q-learn | 0.5879 | 0.1558 |
| Budget Smoothing | 0.3105 | 0.3252 |

Notice that besides *Greedy Product-Limit*, the only other algorithm that competes with the offline opti-mal is our modification to the stochastic knapsack algorithm, *LuekerLearn*. As previously discussed, the convergence of *Q-learn* and *Fixed-Price Search* happens on the time-scale or periods, not auctions. We suspect that with more data, both would converge (as they do in the stochastic setting), albeit *Fixed-Price Search* would converge only to the best fixed-price in hindsight. Finally, as demonstrated in Figure 5, when there is large variation in bidder behavior, *LuekerLearn* might stay bounded away from the offline optimal.

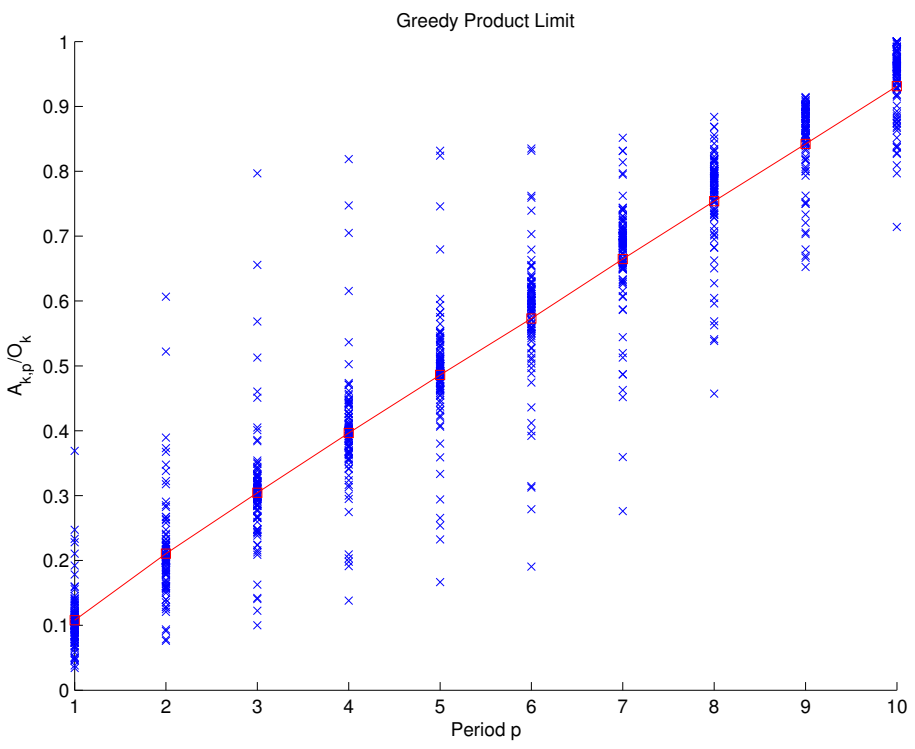

Figure 6: Each scatter point represents $A_{k,p}/O_k$ for a different keyword $k$, at the end of period $p$ displayed on the x-axis. The line is the mean of $A_{k,p}/O_k$ across all $k$ for a fixed period $p$.

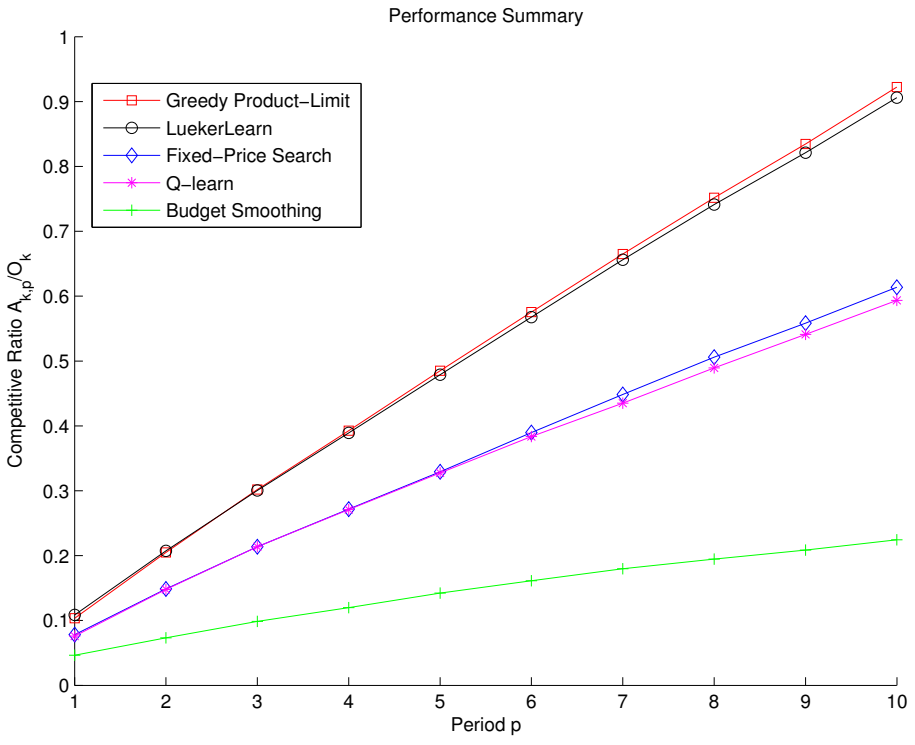

Figure 7: Average of $A_{k,p}/O_k$ across all $k$ for each algorithm.

## 10 Future Work

We conjecture that *Greedy Product-Limit* always converges (rapidly) to the optimal policy in the stochastic setting, and hope to prove so in the future.